\documentstyle[aps,prl,multicol,psfig]{revtex}
\begin{document}
\draft
\title{{\small BioSystems, Vol. 39:117--125, 1996}\\[0.5cm]
The Hypercube Structure of the Genetic Code Explains
Conservative and Non--Conservative Aminoacid Substitutions {\em in
vivo} and {\em in vitro}}
\author{Miguel A. Jim\'enez--Monta\~no\cite{bylinejimm}}
\address{Departamento de F\'{\i}sica y Matem\'aticas, Universidad de
las Am\'ericas,\\
Puebla, Sta. Catarina M\'artir, 72820 Cholula, Puebla, M\'exico}
\author{Carlos R. de la Mora--Bas\'a\~nez\cite{bylinecarlos}}
\address{Direcci\'on General de Investigaciones, Universidad
Veracruzana,\\
Xalapa, Ver. 91000, M\'exico} 
\author{Thorsten P\"oschel\cite{bylinetp}}
\address{Arbeitsgruppe Nichtlineare Dynamik, Universit\"at Potsdam, Am Neuen Palais, D-14415 Potsdam, Germany, and\\ The James Franck
Institute, University of Chicago, 5640 South Ellis Avenue,
Chicago, Illinois 60637} 
\date{\today}
\maketitle
\begin{abstract}
A representation of the genetic code as a six--dimensional Boolean
hypercube is described. This structure is the result of the
hierarchical order of the interaction energies of the bases in
codon--anticodon recognition. In this paper it is applied to study
molecular evolution {\em in vivo} and {\em in vitro}. In the first
case we compared aligned positions in homologous protein sequences and
found two different behaviors: a) There are sites in which the
different amino acids may be explained by one or two ``attractor
nodes'' (coding for the dominating amino acid(s)) and their one--bit
neighbors in the codon hypercube, and b) There are sites in which the
amino acids correspond to codons located in closed paths in the
hypercube. In the second case we studied the ``Sexual
PCR''\footnote{Polymerase Chain Reaction plus DNA shuffling}
experiment described by Stemmer~\cite{Stemmer} and found that the
success of this combination of usual PCR and recombination is in part
due to the Gray code structure of the genetic code.
\end{abstract}

\pacs{PACS numbers: }
\begin{multicols}{2} \narrowtext
\section{Introduction}
The genetic code is the biochemical system for gene expression. It
deals with the translation, or decoding, of information contained in
the primary structure of DNA and RNA molecules into protein
sequences. Therefore the genetic code is both, a physico--chemical
and a communication system. Physically, molecular recognition depends
on the degree of complementarity between the interacting molecular
surfaces (by means of weak interactions); informationally, a
prerequisite to define a code is the concept of distinguishability. It
is the physical indistinguishability of some codon--anticodon
interaction energies that makes the codons synonymous, and the code
degenerate and redundant~\cite{Crick}.

In natural languages~\cite{Harris} as well as in the genetic code the
total redundancy is due to a hierarchy of constraints acting one upon
another. The specific way in which the code departs from randomness
is, by definition, its structure. It is assumed that this structure is
the result of the hierarchical order of the interaction energies of
the bases in codon--anticodon recognition.  The hypercube structure of
the genetic code as currently introduced~\cite{Florida} will be
described and its implications for molecular evolution and test--tube
evolution experiments will be discussed. As we shall see the genetic
code may be represented by a six--dimensional boolean hypercube in
which the codons (actually the code--words; see below) occupy the
vertices (nodes) in such a way that all kinship\footnote{The term
kinship means the relationship between members of the same family.}
neighborhoods are correctly represented. This approach is a particular
application to binary sequences of length six of the general concept
of sequence--space, first introduced in coding theory by
Hamming~\cite{Hamming}.

A code--word is next to six nodes representing codons differing in a
single property. Thus the hypercube simultaneously represents the
whole set of codons and keeps track of which codons are one--bit
neighbors of each other. Different hyperplanes correspond to the four
stages of the evolution of the code according to the Co--evolution
Theory~\cite{Dillon,Wong75,Wong76}. Transitions within
three of the ``columns'' (four--dimensional cubes), consisting of the
codon classes $NGN$, $NAN$, $NCN$, and $NUN$, lead to silent and
conservative amino acid substitutions; while transitions in the same
hyperplane (four--dimensional subspace
belonging to any of the codon classes $ANN$, $CNN$, $GNN$ or $UNN$)
lead to non--conservative substitutions as frequently found in
proteins. The proposed structure demonstrates that in the genetic code
there is a good balance between conservatism and innovation. To
illustrate these results several examples of the non--conservative
variable positions of homologous proteins are discussed. Two different
behaviors were found:
\begin{itemize}
\item[i] There are sites in which the different amino acids 
may be explained by one or two ``attractor nodes'' (coding
for the dominating amino acid(s)) and their one--bit neighbors in the
codon hypercube, and
\item[ii] There are sites in which the amino acids correspond to
codons located in closed paths in the hypercube.
\end{itemize}

Very recently the rapid evolution of a protein {\em in vitro} by DNA
shuffling has been accomplished by Stemmer~\cite{Stemmer}. 

This experiment, called by Smith ``Sexual PCR'', was further discussed
in~\cite{Smith}. Smith recalls that Stemmer investigated the
$beta$--lactamase gene TEM--1 which has a very low activity against
the antibiotic cefoxtamine\footnote{The minimum inhibitory
concentration for Escherichia Coli bacteria carrying TEM--1--bearing
plasmid is only $20~ng~ml^{-1}$.} After three cycles of mutagenesis,
recombination and selection he found the minimum inhibitory
concentration to be $16,000$ times higher than that of the original
clone.

It will be shown that, without exception, the amino--acid
replacements in TEM--1 mutants selected for high resistance to
cefotaxime may be accounted by one bit changes of the corresponding
codons. This shows that the structure of the code permits a very
significant change in function of the coded protein by means of
one--bit changes of some of the codons, provided that these mutations
are integrated in a single polynucleotide by recombination.

\section{Codon--anticodon interaction}
\begin{figure}
\centerline{\psfig{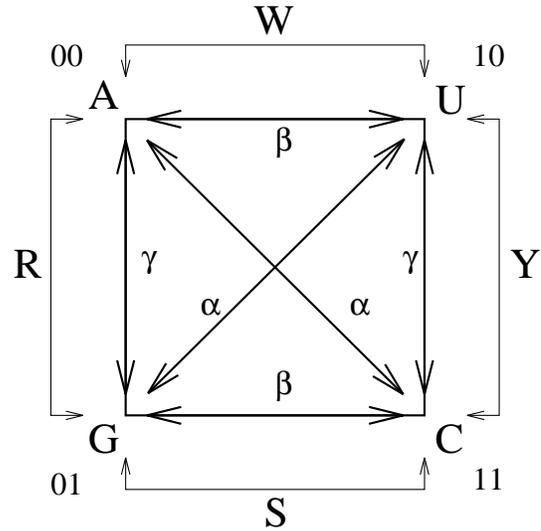}}
\vspace{0.5cm}
\caption{Categorizations of the bases. The categorizations of the
bases according to (i): chemical type ${\cal C}:~\{R,~Y\}$ where
$R:~(A,~G)$ are purines and $Y:~(C,~U)$ are pyrimidines, and (ii)
according to $H$--bonding, ${\cal H}:~\{W,~S\}$, where $W:~(A,~U)$ are
weak and $S:~(C,~G)$ strong bases. The third possible partition into
imino/keto bases is not independent from the former ones and is
irrelevant for the codon--anticodon interaction. The binary
representation of the bases is also shown. The first bit is the
chemical type and the second one the $H$--bonding character. $\alpha$,
$\beta$ and $\gamma$ are the transformations of the bases which form
a Klein--4 group [8,11].
\label{Fig1}
}
\end{figure}

The four bases occurring in DNA (RNA) macromolecules define the
corresponding alphabet $X:\{A,~C,~G,~T\}$ or $X:\{A,~C,~G,~U\}$. Each
base is completely specified by two independent dichotomic
categorizations (Fig.~\ref{Fig1}):
\begin{itemize}
\item[i] according to its chemical type ${\cal C}:\{ R,~Y\}$,
where $R:(A,~G)$ are purines and $Y:(C,~U)$ are pyrimidines and
\item[ii] according to $H$--bonding, ${\cal H}:\{ W,~S\}$, where
$W:~(A,~U)$ are weak and $S:~(C,~G)$ are strong bases.
\end{itemize}

The third possible partition into imino/keto bases is not independent
from the former ones. Denoting by ${\cal C}_i$ the chemical type and
by ${\cal H}_i$ the $H$--bond category of the base $B_i$ at position
$i$ of a codon our basic assumption says that the codon--anticodon
interaction energy obeys the following hierarchical order:
\begin{eqnarray}
&&{\cal C}_2 > {\cal H}_2 > {\cal C}_1 > {\cal H}_1 > {\cal C}_3 > {\cal
H}_3~. \nonumber
\end{eqnarray}

This means, that the most important characteristic determining the
codon--anticodon interaction is the chemical type of the base in the
second position; the next most important characteristic is whether
there is a weak or strong base in this position; then the chemical
type of the first base and so on.

The above assumption goes beyond the early qualitative view that the
optimization between stability and rate, that is always found for
enzyme--substrate interactions, also applies to the codon--anticodon
interaction~\cite{Eigen}. Besides, several authors have suggested that
three bases are needed for effectively binding the adapter to the
messenger. From this it maybe inferred that codon's size determines
a range of codon--anticodon overall interaction strength within which
recognition can occur. Genetic translation rate is limited, among
other things, by codon--anticodon recognition which, in turn, depends
on base--pair lifetimes in a given structural situation. These
life--times are influenced by the nature of the pairs: they are
shorter for $A-T$ than for $G-C$ pairs~\cite{Gueron}.

The bases are represented by the nodes of a 2--cube
(Fig.~\ref{Fig1}). The first attribute is the chemical character and
the second one is the hydrogen--bond character. Extending this
association to base triplets, each codon is associated in a unique way
with a codeword consisting of six attribute values (see Table~1).
\begin{table}[h]
\begin{tabular}{|c c c c c c | c c c | c |}
0 & 0 & 0 & 0 & 1 & 1 & A & A & C & N \\
0 & 0 & 0 & 0 & 1 & 0 & A & A & U & N \\
0 & 0 & 0 & 0 & 0 & 0 & A & A & A & K \\
0 & 0 & 0 & 0 & 0 & 1 & A & A & G & K \\
1 & 0 & 0 & 0 & 0 & 1 & U & A & G & t \\
1 & 0 & 0 & 0 & 0 & 0 & U & A & A & t \\
1 & 0 & 0 & 0 & 1 & 0 & U & A & U & Y \\
1 & 0 & 0 & 0 & 1 & 1 & U & A & C & Y \\
1 & 1 & 0 & 0 & 1 & 1 & C & A & C & H \\
1 & 1 & 0 & 0 & 1 & 0 & C & A & U & H \\
1 & 1 & 0 & 0 & 0 & 0 & C & A & A & Q \\ 
1 & 1 & 0 & 0 & 0 & 1 & C & A & G & Q \\
0 & 1 & 0 & 0 & 0 & 1 & G & A & G & E \\
0 & 1 & 0 & 0 & 0 & 0 & G & A & A & E \\
0 & 1 & 0 & 0 & 1 & 0 & G & A & U & D \\
0 & 1 & 0 & 0 & 1 & 1 & G & A & C & D \\
0 & 1 & 1 & 0 & 1 & 1 & G & U & C & V \\
0 & 1 & 1 & 0 & 1 & 0 & G & U & U & V \\
0 & 1 & 1 & 0 & 0 & 0 & G & U & A & V \\
0 & 1 & 1 & 0 & 0 & 1 & G & U & G & V \\
1 & 1 & 1 & 0 & 0 & 1 & C & U & G & L \\
1 & 1 & 1 & 0 & 0 & 0 & C & U & A & L \\
1 & 1 & 1 & 0 & 1 & 0 & C & U & U & L \\
1 & 1 & 1 & 0 & 1 & 1 & C & U & C & L \\
1 & 0 & 1 & 0 & 1 & 1 & U & U & C & F \\
1 & 0 & 1 & 0 & 1 & 0 & U & U & U & F \\
1 & 0 & 1 & 0 & 0 & 0 & U & U & A & L \\
1 & 0 & 1 & 0 & 0 & 1 & U & U & C & L \\
0 & 0 & 1 & 0 & 0 & 1 & A & U & G & M \\
0 & 0 & 1 & 0 & 0 & 0 & A & U & A & I \\
0 & 0 & 1 & 0 & 1 & 0 & A & U & U & I \\
0 & 0 & 1 & 0 & 1 & 1 & A & U & C & I \\
0 & 0 & 1 & 1 & 1 & 1 & A & C & C & T \\
0 & 0 & 1 & 1 & 1 & 0 & A & C & U & T \\
0 & 0 & 1 & 1 & 0 & 0 & A & C & A & T \\
0 & 0 & 1 & 1 & 0 & 1 & A & C & G & T \\
1 & 0 & 1 & 1 & 0 & 1 & U & C & G & S \\
1 & 0 & 1 & 1 & 0 & 0 & U & C & A & S \\
1 & 0 & 1 & 1 & 1 & 0 & U & C & U & S \\
1 & 0 & 1 & 1 & 1 & 1 & U & C & C & S \\
1 & 1 & 1 & 1 & 1 & 1 & C & C & C & P \\
1 & 1 & 1 & 1 & 1 & 0 & C & C & U & P \\
1 & 1 & 1 & 1 & 0 & 0 & C & C & A & P \\
1 & 1 & 1 & 1 & 0 & 1 & C & C & G & P \\
0 & 1 & 1 & 1 & 0 & 1 & G & C & G & A \\
0 & 1 & 1 & 1 & 0 & 0 & G & C & A & A \\
0 & 1 & 1 & 1 & 1 & 0 & G & C & U & A \\
0 & 1 & 1 & 1 & 1 & 1 & G & C & C & A \\
0 & 1 & 0 & 1 & 1 & 1 & G & G & C & G \\
0 & 1 & 0 & 1 & 1 & 0 & G & G & U & G \\
0 & 1 & 0 & 1 & 0 & 0 & G & G & A & G \\
0 & 1 & 0 & 1 & 0 & 1 & G & G & G & G \\
1 & 1 & 0 & 1 & 0 & 1 & C & G & G & R \\
1 & 1 & 0 & 1 & 0 & 0 & C & G & A & R \\
1 & 1 & 0 & 1 & 1 & 0 & C & G & U & R \\
1 & 1 & 0 & 1 & 1 & 1 & C & G & C & R \\
1 & 0 & 0 & 1 & 1 & 1 & U & G & C & C \\
1 & 0 & 0 & 1 & 1 & 0 & U & G & U & C \\
1 & 0 & 0 & 1 & 0 & 0 & U & G & A & t \\
1 & 0 & 0 & 1 & 0 & 1 & U & G & C & W \\
0 & 0 & 0 & 1 & 0 & 1 & A & G & G & R \\
0 & 0 & 0 & 1 & 0 & 0 & A & G & A & R \\
0 & 0 & 0 & 1 & 1 & 0 & A & G & U & S \\
0 & 0 & 0 & 1 & 1 & 1 & A & G & C & S 
\end{tabular}
\caption{Gray code representation of the genetic code. In the first
and fourth blocks the six--dimensional vectors
(code--words) are shown. In the second and fifth blocks appear the
corresponding codons. Finally, in the third and sixth columns the
amino acids in single letter notation. The first two digits correspond
to the first base, the following two to the second base and the last
two to the last base, according to the binary codification of the
bases of Fig.~1.
}
\end{table}

\begin{figure}
\centerline{\psfig{figure=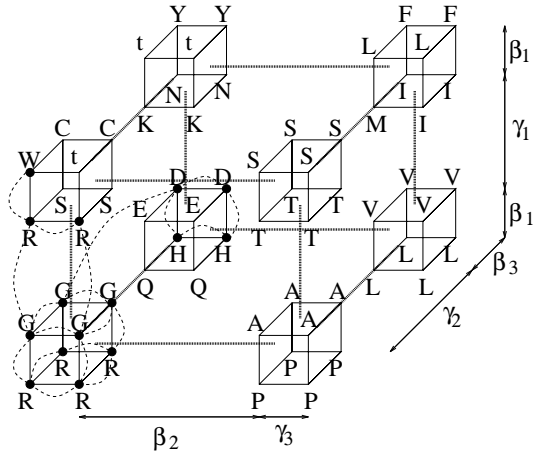,width=7cm,angle=270}}
\vspace{0.5cm}
\caption{The six--dimensional hypercube. Each node is labeled with the
corresponding amino acid in the single letter notation or terminator
symbol. The fat short dashed lines represent a complex connection
between two (three--dimensional) cubes. Such a line represents 8 edges
each, connecting the corresponding nodes of two neighbored
three--dimensional cubes (see fig.~\ref{4d}). The cluster of amino
acids of the first example discussed in the text is displayed by fat
points at the corresponding nodes and dashed thin curved lines for the
edges.}
\label{Fig2a}
\end{figure}

In some of the hypercube directions single feature codon changes
(one--bit code--word changes) produce synonymous or conservative amino
acid substitutions in the corresponding protein (when the 
transitions occur in three of the 4--cubes displayed as
``columns'' in Figs.~\ref{Fig2a} and \ref{Fig2b}); while in other
directions lead to context dependent replacements which, in general,
conserve only certain physical properties. However, if these
properties are the only relevant ones in the given context, the
substitution has little effect on the protein structure as well. These
low--constraint sites facilitate evolution because they allow the
transit between hypercube columns belonging to amino acids with very
different physico--chemical properties (e.g. hydrophobic and
hydrophilic amino acids, respectively).

\begin{figure}
\centerline{\psfig{figure=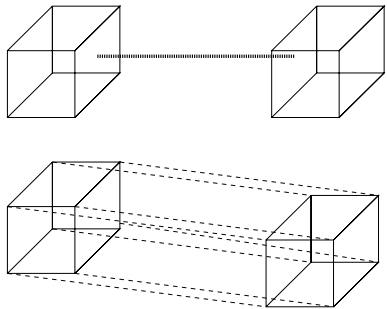,width=5cm,angle=270}}
\caption{Each of the fat short dashed lines represent 8 edges,
connecting the corresponding nodes of two three--dimensional cubes.
The figure shows a four--dimensional cube using the symbolic fat drawn
link (top) and the same cube using standard representation.}
\label{4d}
\end{figure}

\begin{figure}
\centerline{\psfig{figure=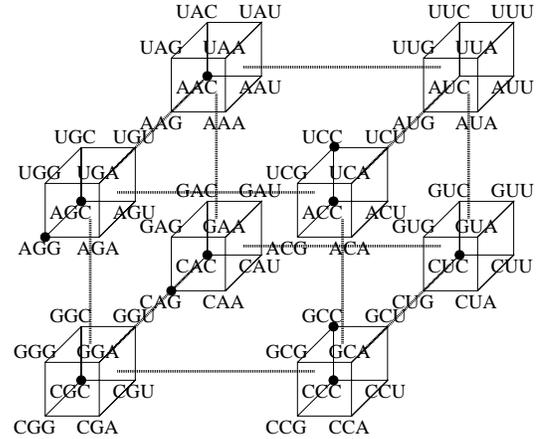,width=7cm,angle=270}}
\vspace{0.5cm}
\caption{The hypercube representation of the genetic code. Each node
represents a code--word (six--dimensional vector) of attribute
values. However, for clarity of interpretation, the nodes are labeled
with the corresponding codons (See Table~1 for the
assignment of codons to vectors). The nodes and links mentioned in
second example discussed in the text are shown. The edges connect:
$AGG \leftrightarrow AGC$,
$AGC \leftrightarrow ACC$,
$AGC \leftrightarrow AAC$,
$UCC \leftrightarrow ACC$,
$ACC \leftrightarrow GCC$,
$GCC \leftrightarrow CCC$,
$CGC \leftrightarrow CAC$,
$CAC \leftrightarrow CAG$,
$CAC \leftrightarrow CUC$,
$CAC \leftrightarrow GAC$}
\label{Fig2b}
\end{figure}

\section{Gray code structure of the genetic code}
An $n$--dimensional hypercube, denoted by $Q_n$, consists of $2^n$
nodes each addressed by a unique $n$--bit identification number. A
link exists between two nodes of $Q_n$ if and only if their node
addresses differ in exactly one bit position. A link is said to be
along dimension $i$ if it connects two nodes which addresses differ to
as the $i$th bit (where the least significant bit is referred to as
the $0$th bit). $Q_6$ is illustrated in Fig.~\ref{Fig2b}. Two nodes in
a hypercube are said to be adjacent if there is a link between
them. The (Hamming) distance between any two cube nodes is the number
of bits differing in their addresses. The number of 
transitions needed to reach a node from another node equals the
distance between the two nodes. A $d$--dimensional sub-cube in $Q_n$
involves $2^d$ nodes which addresses belong to a sequence of $n$
symbols $\{0,~1,~*\}$ in which exactly $d$ of them are of the symbol
$*$ (i.e. the don't care symbol which value can be $0$ or $1$).

The idea to propose a Gray Code representation of the Genetic Code
goes back to Swanson~\cite{Swanson} where this concept is explained
in detail (see also~\cite{Jimenez}). However, a great number of
different Gray Codes can be associated to the Genetic Code depending
on the order of importance of the bits in a code--word. In
Table~1 our chosen Gray Code is displayed. It is
constructed according to our main hypothesis
\begin{eqnarray}
&&{\cal C}_2 > {\cal H}_2 > {\cal C}_1 > {\cal H}_1 > {\cal C}_3 > {\cal
H}_3 ~.\nonumber 
\end{eqnarray}

For example, the first two lines of the table differ in the last bit
corresponding to ${\cal H}_3$; which is the least significant bit; the
second and the third lines differ in the next least significant bit,
i.e. ${\cal C}_3$, and so forth.

\section{The structure of codon doublets}
This section is more mathematical than the rest of the paper. It is
not essential for the understanding of the rest of the paper.

\begin{figure}
\centerline{\psfig{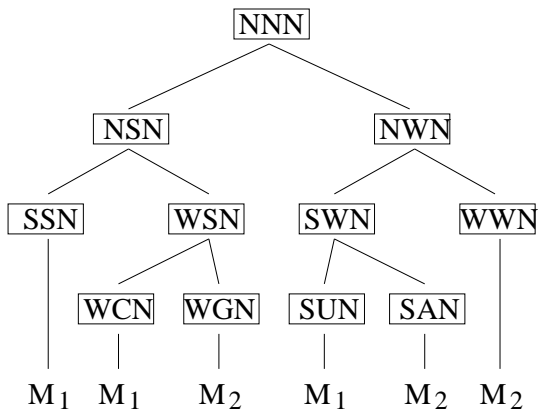}}
\vspace{0.5cm}
\caption{Decision--tree of codon categories and redundancy
distribution. The leaves are the sets of four--fold ($M_1$) and less
than four--fold ($M_2$) degenerate $B_1B_2$ doublets.}
\label{Fig3}
\end{figure}

In a pioneering paper Danckwerts and Neubert~\cite{Danckwerts}
discussed the symmetries of the sixteen $B_1B_2$ codon doublets in
terms of the Klein--4 group of base transformations. Here their result
will be recast in a form of a decision--tree (Fig.~\ref{Fig3}) and
their analysis will be extended to the $B_2B_3$ doublets. They found
the following structure for the set $M$ of $B_1B_2$ doublets:

Starting from $Ac$ generate the set:
\begin{eqnarray}
M_0 &=& \left\{\left[\left(1,1\right) \cup \left(\alpha,1\right) \cup
        \left(\alpha,\beta\right) \cup \left(\alpha,\gamma\right)
        \right] AC \right\}\nonumber\\ 
    &=& \left\{ AC, CC, CG, CU \right\} \nonumber\\
M_1 &=& \left[ \left(1,1\right) \cup \left(\beta,1\right) \right]
        M_0\nonumber\\
M_2 &=& \left(\alpha,\alpha\right) M_1\nonumber
\end{eqnarray}
The sets $M_1$ and $M_2$ consist of four--fold and less than
four--fold degenerate doublets, respectively.

The set $M$ can be expressed as:
\begin{eqnarray}
M &=& \left[\left(1,1\right)\cup \left(b,1\right)\right]
\left[\left(1,1\right) \cup \left(\alpha, \alpha\right)\right]
M_0\nonumber 
\end{eqnarray}
Where the base exchange operators $\alpha, \beta,\gamma$ are defined
in Fig.~\ref{Fig1}. 

They showed that: ``a) $M_1$ and $M_2$ are invariant by operating with
$(\beta, 1)$ on $B_1$, but no operation on $B_2$ leaves $M_1$ or $M_2$
invariant. Thus $B_2$ carries more information
than $B_1$ and $B_2$ is therefore more
important for the stability of $M_1$ and $M_2$ than $B_1$\dots $A$
change of $B_1$ with respect to its hydrogen bond property does not
change the resulting amino acids if all doublets of either $M_1$ or
$M_2$ are affected.

Reversing supposition and conclusion, $M_1$ and $M_2$ may be defined
as those doublet sets of 8 elements which are invariant under the
$(\beta,1)$--transformation. Then experience shows that $M_1$ and
$M_2$ are fourfold and less than fourfold degenerate respectively.''

Thus the third base degeneracy of a codon does not depend on the
exact base $B_1$, but only on its $H$--bond property (weak or strong).

The above results can be simply visualized as a decision--tree
(Fig.~\ref{Fig3}). It can be seen from this figure that the redundancy
of a codon is determined only by the $H$--bond character of $B_1$ and
$B_2$: $SSN$ codons (with 6 $H$--bonds in $B_1B_2$) belong to $M_1$
while $WWN$ codons (with 4 $H$--bonds in $B_1B_2$) belong to
$M_2$. However, for codons $WSN$ and $SWN$ (with 5 $H$--bonds in
$B_1B_2$) it is not possible to decide unless one has more information
about the second base: $WCN$ and $SUN$ belong to $M_1$ while $WGN$ and
$SAN$ belong to $M_2$. In all cases at most three attributes are
necessary to determine the redundancy of a codon up to this point. Of
course the non--degenerate codons ($UAG$ for Methionine and $UGG$ for
Tryptophan) will require the specification of the six attributes.

From the decision rules obtained from Fig.~\ref{Fig3} it is clear that
there are branches where the refinement procedure cannot continue (the
branches which end in $M_1$) because no matter which base occupies
the third codon position the degeneracy cannot be lifted. This imposes
a limit to the maximum number of amino acids which can be incorporated
to the code without recurring to a ``frozen accident'' hypothesis. Our
proposal generalizes the ``2--out--of--3'' hypothesis of
Lagerkvist~\cite{Lagerkvist} which refers only to codons in the $SSN$
class.

\begin{figure}
\centerline{\psfig{figure=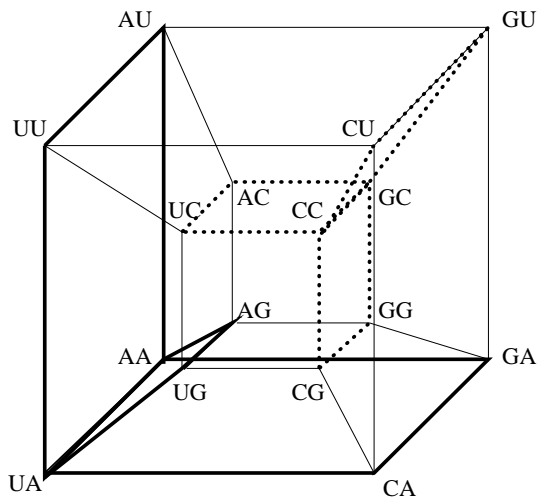,width=7cm,angle=270}}
\vspace{0.5cm}
\caption{The four--dimensional hypercube
representation of the sets $M_1$ (dotted) and $M_2$ (fat).}
\label{Fig4a}
\end{figure}

The sixteen $B_1B_2$ doublets can be represented as the vertices of a
four--dimensional hypercube. Figure~\ref{Fig4a} shows that the
sets $M_1$ and $M_2$ are located in compact regions. Notice that this
figure differs from the one introduced by Bertman and
Jungck~\cite{Bertman} who considered as basic transformations
$\alpha$ and $\beta$, instead of $\beta$ and $\gamma$ as we did. Since
the operator $\alpha$ changes two bits we do not consider it as basic.

\begin{figure}
\centerline{\psfig{figure=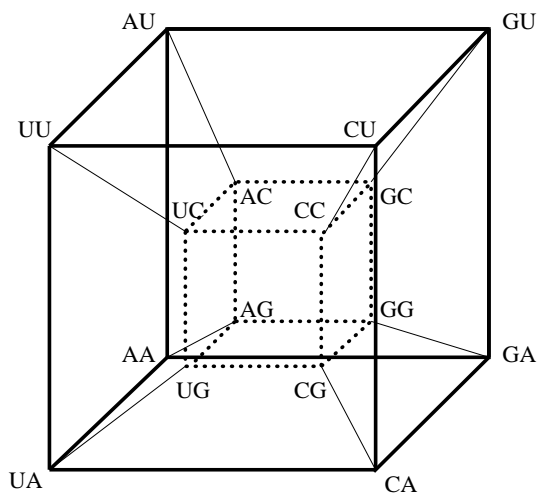,width=7cm,angle=270}}
\vspace{0.5cm}
\caption{The corresponding to fig.~\ref{Fig4a} hypercube of the sets
$M_1^\prime$ and $M_2^\prime$ (see text). Notice that in both cases
each set is located in a compact region.}
\label{Fig4b}
\end{figure}

Let's consider now the structure of the set $M^\prime$ of $B_2B_3$
doublets: Exactly as before, define the sets
\begin{eqnarray}
M_0^\prime &=& \left\{NC\right\} \nonumber\\
M_1^\prime &=& \left[\left(1,1\right) \cup \left(1,\beta\right)\right]
               M_0^\prime \nonumber\\ 
M_2^\prime &=& \left(\alpha,\alpha\right) M_1^\prime 
               \mbox{ (alternatively } 
M_1^\prime =\left(\alpha,\alpha\right) M_2^\prime \mbox{ )}~,\nonumber
\end{eqnarray}
where $M_1^\prime$ consists of the doublets $B_2B_3$ ending in a
strong base (NS) and $M_2^\prime$ of the doublets ending in a weak
base (NW).

Then
\begin{eqnarray}
M^\prime &=& M_1^\prime \cup M_2^\prime \nonumber
\end{eqnarray}
can be expressed as
\begin{eqnarray}
M^\prime &=& \left[\left(1,1\right) \cup
\left(1,\beta\right)\right]\left[\left(1,1\right) \cup
\left(\alpha,\alpha\right)\right] 
M_0^\prime~.\nonumber
\end{eqnarray}

Notice that the operator acting on $M_0^\prime$ has the same
functional form as the operator acting on $M_0$ above, except that
$\beta$ acts as the third base instead of the first.

The sets $M_1^\prime$ and $M_2^\prime$ are invariant under the
$(1,\beta)$--transformations. Then experience shows that the 32 codons
in the class $NB_2B_3$, with $B_2B_3$ in $M_1^\prime$ or $M_2^\prime$
constitute a complete code codifying for the 20 amino acids and
terminator signal (stop--codon), if allowance is made for deviating
codon--assignments found in Mitochondria~\cite{Jukes}. For the codons
in $M_1^\prime$ this is true in the universal code; for codons in
$M_2^\prime$ $AUA$ should codify for $M$ instead of $I$ and $UGA$ for
$W$ instead of stop signal. Both changes have been observed in
Mitochondria. This more symmetric code has been considered more similar
to an archetypal code than the universal code~\cite{Jukes}. Only after
the last attribute $H_3$ was introduced the universal code was
obtained, with the split of $AUR$ into $AUA$ ($I$) and $AUG$ ($M$)
and $UGR$ into $UGG$ ($W$) and $UGA$ ($t$).

It has been speculated that primordial genes could be included in a
$0.55~kb$ open reading frame~\cite{Naora}. The same authors calculated
that with two stop codons this open reading frames would have appeared
too frequently. From the present view the assignment of $UGA$ to a
stop codon was a late event that optimized this frequency (this
interpretation differs from the one proposed in~\cite{Naora,Brentani}
where a primordial code with three stop codons) is assumed. Other
deviations of the universal code most likely also occurred in the last
stages of the code's evolution.

In the same way as before the sixteen $B_2B_3$ doublets can be
represented as the vertices of a four--dimensional hypercube
(Fig.~\ref{Fig4b}). The sets $M_1^\prime$ and $M_2^\prime$ are also
located in compact regions. Codons with $B_2B_3$ in $M_1^\prime$ are
frequently used in eukaryotes. In contrary, codons with $B_2B_3$ in
$M_2^\prime$ are frequently used in prokaryots. The described
structure of the code allows a modulation of the codon--anticodon
interaction energy~\cite{Grosjean}.

\section{Examples}
Besides the results mentioned in the last section which refer to
codon doublets, to further illustrate the significance of proposed
approach, we are going to consider several examples of molecular
evolution.

The first example (Fig.~\ref{Fig2a}) refers to the alignment studied
using the method of hierarchical analysis of residue
conservation by Livingstone and Barton (Fig.~2
in~\cite{Livingstone}). In position 11 appear the following amino
acids $R$, $W$, $H$, $G$, $D$, which according to their approach have
no properties in common. In Fig.~\ref{Fig2a} this cluster of amino
acids is shown. By looking at the Atlas of amino acid
properties~\cite{Nakai} we see that from the properties proposed by
Grantham~\cite{Grantham} (composition, polarity and volume) apparently
the only requirement for the amino acids at this site is to maintain a
certain degree of polarity. From this observation we may conclude that
most probably it is an external site.  Simply by looking at such a
diverse set of amino acids one can hardly realize that they have
clustered codons. This clustering facilitates the occurrence of
mutations that in the course of evolution were fixed, in view of the
low physico--chemical requirements at the site.

\begin{figure}
\centerline{\psfig{figure=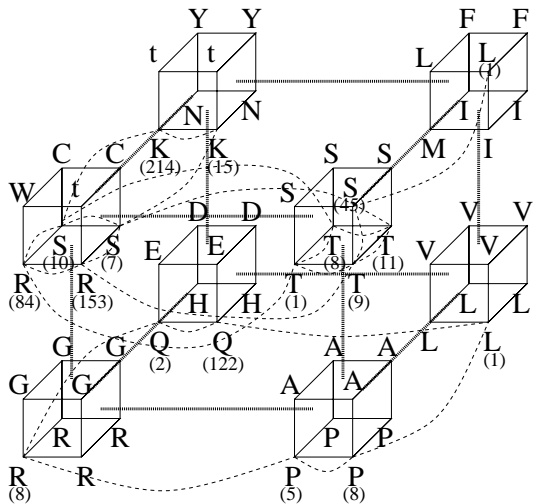,width=7cm,angle=270}}
\vspace{0.5cm}
\caption{The amino acid hypercube with the amino acids at position 18
of the variable region of kappa light chain displayed. The number
after the amino acid symbol is the number of times the amino acid
occurs in the alignment in~[24].
}
\label{Fig5}
\end{figure}

As a second example (Fig.~\ref{Fig2b}) let us consider site 33 of the
alignment of 67 $SH2$ domains, Fig.~6 of~\cite{Gueron}. We can see
from Fig.~\ref{Fig2b} that the cluster around the codon $CAC$ ($H$)
explains, by one--bit changes, the amino acids $R$, $Q$, $L$, $H$,
$D$. Furthermore, a second cluster around the codon $AGC$ ($S$)
explains the aminoacids $R$, $N$, $S$, $T$. Finally, a silent change
from $AGC$ ($S$) to $UCC$ ($S$) accounts for the minor appearance of
the small, neutral amino acids, $A$, $T$, $P$. In a similar way the
variation of the hyper--variable region of immunoglobulin kappa light
FR1 at position 18 can be explained (Fig.~\ref{Fig5}). The number
after the amino acid symbol in Fig.~\ref{Fig5} is the number of times
the amino acid occurs in the alignment in~\cite{Kabat}.
As a third example consider the residue frequencies in 226 globins
displayed in Table~3 of the paper by Bashford et
al.~\cite{Bashford}. From this table we find that there are variable
positions in which one or two residues predominately occur and the
rest are only marginally represented and others in which the
frequencies are more evenly distributed among the amino acids. As it
can be easily shown, the first class of positions may be associated,
at the codon level, with one (or two) attractor node(s) and its
one--bit neighbors.  The second one can be associated with closed
trajectories in the hypercube. The corresponding figures are not
included because of lack of space.

Finally, let us discuss the ``sexual PCR'' experiment. In the paper by
Smith~\cite{Smith} a table is displayed showing the positions in the
TEM--1 gene where mutations occur, together with the substitutions
found in the variant genes ST--1, ST--2 and ST--4 which show increased
resistance to cefotaxime. We refer to the mentioned paper for further
details. Locating these mutations in the hypercube (Fig.~\ref{Fig2b})
one can easily convince oneself that all mutations may be accounted by
one--bit changes at the codon level. Therefore only six codons (four
or five aminoacids) are searched in each mutation and not 19
alternatives. This finding helps to explain why this {\em in vitro}
realization of a ``genetic algorithm'' was so successful.

It is well known in the field of Genetic Algorithms that a proper
encoding is crucial to the success of an algorithm. Furthermore
in~\cite{Caruna} it is shown the superiority of Gray coding over
binary coding for the performance of a genetic algorithm. As it was
shown above the structure of the genetic code is precisely the
structure of a Gray code. Therefore it is our claim that this is one
of the reasons why very efficient variants were found after very few
rounds of recombination. Most probably other reasons are: the initial
population was not random, but consisted of selected sequences and
these sequences were very similar among themselves. This explanation
of the results of Stemmer's experiment differs from the explanation
advanced by Smith~\cite{Smith}.

\section{Concluding Remarks}
The present approach goes beyond the usual analyses in terms of single
base changes, because it takes into account the two characters of each
base and therefore it represents one--bit changes. Besides, the base
position within the codon is also considered. The fact that single bit
mutations occur frequently is expected from probabilistic
arguments. However, one could not expect, a priori, that a cluster of
mutations would correspond, at the amino acid level, to a cluster of
amino acids fixed by natural selection. We have found that this
situation presents itself for many positions of homologous protein
sequences of many different families (results not included). The
structure of the code facilitates evolution: the variation found at
the variable positions of proteins do not corresponds to random jumps
at the codon level, but to well defined regions of the
hypercube. Finally, the Gray code structure of the genetic code helps
to explain the success of ``Sexual PCR'' experiments.

\acknowledgments
This work received economical support from: Proyecto CONACyT
No. 1932--E9211. We thank professors Werner Ebeling and Michael Conrad
for encouraging comments.

\end{multicols}

\begin{references}
\bibitem[*]{bylinejimm} e--mail: jimm@udlapvms.pue.udlap.mx
\bibitem[+]{bylinecarlos} e--mail: delamora@speedy.coacade.uv.mx
\bibitem[**]{bylinetp} e--mail: thorsten@hlrsun.hlrz.kfa--juelich.de\\
http://summa.physik.hu--berlin.de:80/~thorsten/

\bibitem{Stemmer} W.~P.~C.~Stemmer, Rapid evolution of a protein in
vitro by DNA shuffling, {\em Nature}, {\bf 370}, 389--391 (1994).

\bibitem{Crick} F.~Crick, Codon--anticodon pairing: the wobble
hypothesis. {\em J.~Mol.~Biol.}, {\bf 19} 548--555 (1966).

\bibitem{Harris} Z.~Harris, {\em Language and Information.} Columbia
University Press (New York, 1988).

\bibitem{Florida} M.~A.~Jim\'enez--Monta\~no, C.~R.~de la
Mora--Bas\'a\~nez, and T.~P\"oschel, On the Hypercube Structure of the
genetic Code {\em preprint} (1994).

\bibitem{Hamming} R.~W.~Hamming, {\em Bell Syst. Tech. J.}, {\bf 29},
147--160 (1950); W.~Ebeling and R.~Feistel, {\em Physik der
Selbstorganisation und Evolution.} Akademie--Verlag (Berlin, 1982);
W.~Ebeling, R.~Feistel, and M.~A.~Jim\'enez--Monta\~no, On the theory
of stochastic replication and evolution of molecular sequences, {\em
Rostocker Phy\-si\-ka\-li\-sche Manuskripte}, {\bf 2}, 105--127 (1977).

\bibitem{Dillon} L.~Dillon, {\em The genetic mechanism and the origin of
life.} Plenum Press (1978).

\bibitem{Wong75} J.~Wong, A co--evolution theory of the genetic
code. {\em Proc.~Nat.~Acad.~Sci.~USA}, {\bf 72}, 1909--1912 (1975).

\bibitem{Wong76} J.~Wong, The evolution of a universal genetic code,
{\em Proc.~Nat.~Acad.~Sci.~USA}, {\bf 73}, 2336--2340 (1976).

\bibitem{Smith} G.~P.~Smith, The progeny of sexual PCR, {\em Nature},
{\bf 370}, 324--325 (1994).

\bibitem{Eigen} M.~Eigen, Selforganization of matter and the evolution
of biological macromolecules, {\em Naturwiss.}, {\bf 58}, 465--523
(1971).

\bibitem{Gueron} M.~Gu\'eron, E.~Charretier, M.~Kochoyan, and
J.~L.~Leroy, Applications of imino proton exchange to nucleic acid
kinetics and structures, in: {\em Frontiers of NMR in Molecular
Biology.}, 225--238 (Alan R.~liss Inc., 1990).

\bibitem{Swanson} R.~Swanson, A unifying concept for the amino acid
code, {\em Bull.~Math.~Biol.}, {\bf 46}, 187--204 (1984).

\bibitem{Jimenez} M.~A.~Jim\'enez--Monta\~no, On the syntactic
structure and redundancy distribution of the genetic code, {\em
BioSystems}, {\bf 32}, 11--23 (1994).

\bibitem{Danckwerts} H.~J.~Danckwerts and D.~Neubert, symmetries of
genetic code--doublets, {J.~Mol.~Evol.}, {\bf 5}, 327--332 (1975).

\bibitem{Lagerkvist} U.~Lagerkvist, ``Two out of three'': an
alternative method for codon reading, {\em Proc.~Nat.~Acad.~Sci.~USA},
{\bf 75}, 1759--1762 (1978); Unorthodox codon reading and the
evolution of the genetic code, {\em Cell}, {\bf 23}, 305--306 (1981).

\bibitem{Bertman} M.~O.~Bertman and J.~R.~Jungck, Group graph of the
genetic code, {\em J.~Heredity}, {\bf 70}, 379--384 (1979).

\bibitem{Jukes} T.~H.~Jukes, Evolution of the amino acid code, in:
M.~Nei and R.~K.~Koehn (eds.), {\em Evolution of Genes and Proteins},
Sinauer Associates Inc, 191--207 (Sunderland, Mass, 1983).

\bibitem{Naora} H.~Naora, K.~Miyahara, and R.~N.~Curnow (1987). Origin
of non-coding DNA sequences: molecular fossils of genome evolution,
{\em Proc.~Natl.~Acad.~Sci.~USA}, {\bf 84}, 6195--6199 (1987).

\bibitem{Brentani} R.~R.~Brentani, Complementary hydropathy and the
evolution of interacting polypeptides, {\em J.~Mol.~Evol.}, {\bf 31},
1239--243 (1990). 

\bibitem{Grosjean} H.~Grosjean, D.~Sankoff, W.~Min Jou, W.~Fiers, and
R.~J.~Cedergren, Bacteriophage MS2 RNA: A correlation between the
stability of the codon: anticodon interaction and the choice of code
words, {\em J.~Mol.~Evol.}, {\bf 12}, 113--119 (1978); distribution of
the genetic code, {\em BioSystems}, {\bf 32}, 11--23. 

\bibitem{Livingstone} C.~D.~Livingstone and G.~J.~Barton, Protein
sequence alignments: a strategy for the hierarchical analysis of
residue conservation, {\em CABIOS}, {\bf 9}, 745--756 (1993).

\bibitem{Nakai} K.~Nakai, A.~Kidera, and M.~Kanehisa, Cluster analysis
of amino acid indices for prediction of protein structure and
function, {\em Prot.~Eng.}, {\bf 2}, 93--100 (1988).

\bibitem{Grantham} R.~Grantham, Amino acid difference formula to help
explain protein evolution, {\em Science}, {\bf 185}, 862--864 (1974).

\bibitem{Kabat} Kabat et al. {\em Sequences of proteins of
immunological interest}, 5th ed. NIH (Bethesda, 1991).

\bibitem{Bashford} D.~Bashford, C.~Chothia, and A.~M.~Lesk,
Determinants of a protein fold, {\em J.~Mol.~Biol.}, {\bf 196},
199--216 (1987).

\bibitem{Caruna} R.~A.~Caruana and J.~D.~Schaffer, Representation and
hidden bias: Gray vs. binary coding for genetic algorithms, in:
J.~Laird (ed.), {\em Proceedings of the Fifth International Conference
on Machine Learning}, Morgan Kauffman Publ. Inc., 153--161 (San Mateo,
1988).
\end{references}
\end{document}